\documentclass[journal,twoside,web]{IEEEcolor}
\usepackage{generic}
\usepackage{cite}
\usepackage{amsmath,amssymb,amsfonts}
\usepackage{algorithmic}
\usepackage{graphicx}
\usepackage{textcase}

\usepackage{balance}
\usepackage{url}
\usepackage{hyperref}
\def\BibTeX{{\rm B\kern-.05em{\sc i\kern-.025em b}\kern-.08em
    T\kern-.1667em\lower.7ex\hbox{E}\kern-.125emX}}
\begin{document}

\title{ Transistors based on Novel 2-D Monolayer Semiconductors Bi$_2$O$_2$Se, InSe, and MoSi$_2$N$_4$ for Enhanced Logic Density Scaling 

}
%%%%%%%%%%%
\author{Keshari Nandan, \IEEEmembership{Member, IEEE}, Ateeb Naseer, \IEEEmembership{Graduate Student Member, IEEE}, Amit Agarwal, Somnath Bhowmick, Yogesh S. Chauhan, \IEEEmembership{Fellow, IEEE}
\thanks{K. Nandan was with Department of Electrical Engineering, IIT Kanpur, Kanpur
	208016, India. He is now with the Department of Electrical and Computer Engineering,
	University of Minnesota, Minneapolis MN 55455, USA (e-mail: knandan@umn.edu).}
	\thanks{A. Naseer and Y. S. Chauhan are with Department of Electrical Engineering, IIT Kanpur, Kanpur 208016, India.(e-mail: chauhan@iitk.ac.in).}
	\thanks{A. Agarwal is with Department of Physics, IIT Kanpur, Kanpur 208016, India.}
\thanks{S. Bhowmick is with Department of Materials Science and Engineering, IIT Kanpur, Kanpur 208016, India.}}
\maketitle

\begin{abstract}
	Making ultra-short gate-length transistors significantly contributes to scaling the contacted gate pitch. This, in turn, plays a vital role in achieving smaller standard logic cells for enhanced logic density scaling.
As we push the boundaries of miniaturization, it is intriguing to consider that the ultimate limit of contacted gate pitch could be reached with remarkable 1 nm gate-length transistors. Here, we identify InSe, Bi$_2$O$_2$Se, and MoSi$_2$N$_4$ as potential two-dimensional semiconductors for 1 nm transistors with low contact resistance and outstanding interface properties. We employ a fully self-consistent ballistic quantum transport model starting from first-principle calculations. Our simulations show that the interplay between electrostatics and quantum tunneling influences the performance of these devices over the device design space. MoSi$_2$N$_4$ channels have the best immunity to quantum tunneling, and Bi$_2$O$_2$Se channel devices have the best electrostatics. We show that for a channel length of 12 nm, all the devices can deliver $I_{ON}/I_{OFF}$ $>$ $10^3$, suitable for electronic applications, and Bi$_2$O$_2$Se is the best-performing channel material. 
\end{abstract}


\section{Introduction}
\label{sec:introduction}
FETs based on 2-D channel materials are more immune to short-channel effects (SCEs) and offer better device scaling \cite{zhuo2023modifying, 10.3389/felec.2023.1277927}. In ultra-thin-body devices, they show superior gate control and better mobility than bulk semiconductors such as Si, Ge, and III-V. In recent years, experimental 2-D FETs have shown ideal perormance. In addition, one can integrate a 2-D channel with a 1-D metallic nanowire, nanotube, or 2-D graphene as the gate. This integration facilitates 1 nm and sub-1 nm gate lengths in 2-D FETs \cite{cao2016prospects, Desai99, wu2022vertical, xiao2022locally, liu2018mos2, jolie20241d}.
Considering the interface quality of the gate stack, metal contact with 2-D semiconductors, and quality of grown channel,
we study a few potential contenders for the channel material in ultra-short 2-D FETs. They are monolayers of MoSi$_2$N$_4$, Bi$_2$O$_2$Se, and InSe.

MoSi$_2$N$_4$ was synthesized in 2020, showing excellent physical, mechanical, thermal, electronic, and metal contact properties \cite{Hong, Wang2021, 9646230, PhysRevApplied.19.064058}. It outperforms most other 2-D semiconductors and shows excellent ohmic contact with CMOS-compatible metals \cite{Wang2021}.
Bi$_2$O$_2$Se is a ternary semiconducting compound with outstanding stability and a moderate bandgap \cite{li20212d, quhe2019high}. High-$k$ native crystalline oxide dielectric, Bi$_2$SeO$_5$, can be formed by layer-by-layer oxidation of Bi$_2$O$_2$Se. At wafer-scale, Bi$_2$O$_2$Se FETs can be fabricated using this native oxide as gate oxide with EOT below sub-0.5 nm \cite{zhang2022single}. The performance of wafer-scale Bi$_2$O$_2$Se based FinFETs with Bi$_2$SeO$_5$ as gate insulator is competitive to Si and Ge based FinFETs. These Bi$_2$O$_2$Se FinFETs show low contact resistance, $R_\mathrm{C}$ $\sim 470~\Omega~\mu$m \cite{tan20232d}. InSe FETs have shown ideal performance close to the ballistic limit down to 10 nm gate length with a power supply voltage of 0.5 V and  low contact resistance, $R_\mathrm{C}$ $\sim$ 65 $\Omega~\mu$m \cite{jiang2023ballistic}. Wafer-scale InSe meets the standard for electronic-grade semiconductor material \cite{song2023wafer}. These are ideal atomic thin channel materials for next-generation logic design and monolithic three-dimensional (M3D) integration with state-of-the-art Si devices.

Here, we explore the potential of $\sim$ 1 nm gate-length transistors using novel 2-D semiconducting monolayers, highlighting their excellent performance characteristics and role in advancing electronic device miniaturization. Our research highlights the role of geometrical and material parameters—such as the nanowire gate, spacer, gate dielectric, and channel length  —within integrated 2-D channel FETs featuring metallic nanowires.
	Considering the favorable figures for 2-D semiconductors for the transistor's channel, we choose monolayers of MoSi$_2$N$_4$, Bi$_2$O$_2$Se, and InSe as channels in our research.
	The possibility of integrating a 2-D channel with a 1-D metallic nanowire as a gate alleviates the lithography limitation on gate pattering \cite{cao2016prospects, Desai99, wu2022vertical, xiao2022locally}. This configuration allows one to realize the scaling potential of 2-D FETs, thanks to their natural atomic-scale dimensions. The key finding of our research are following-

\begin{figure}[!t]
	\centering
	\includegraphics[width = 0.49\textwidth]{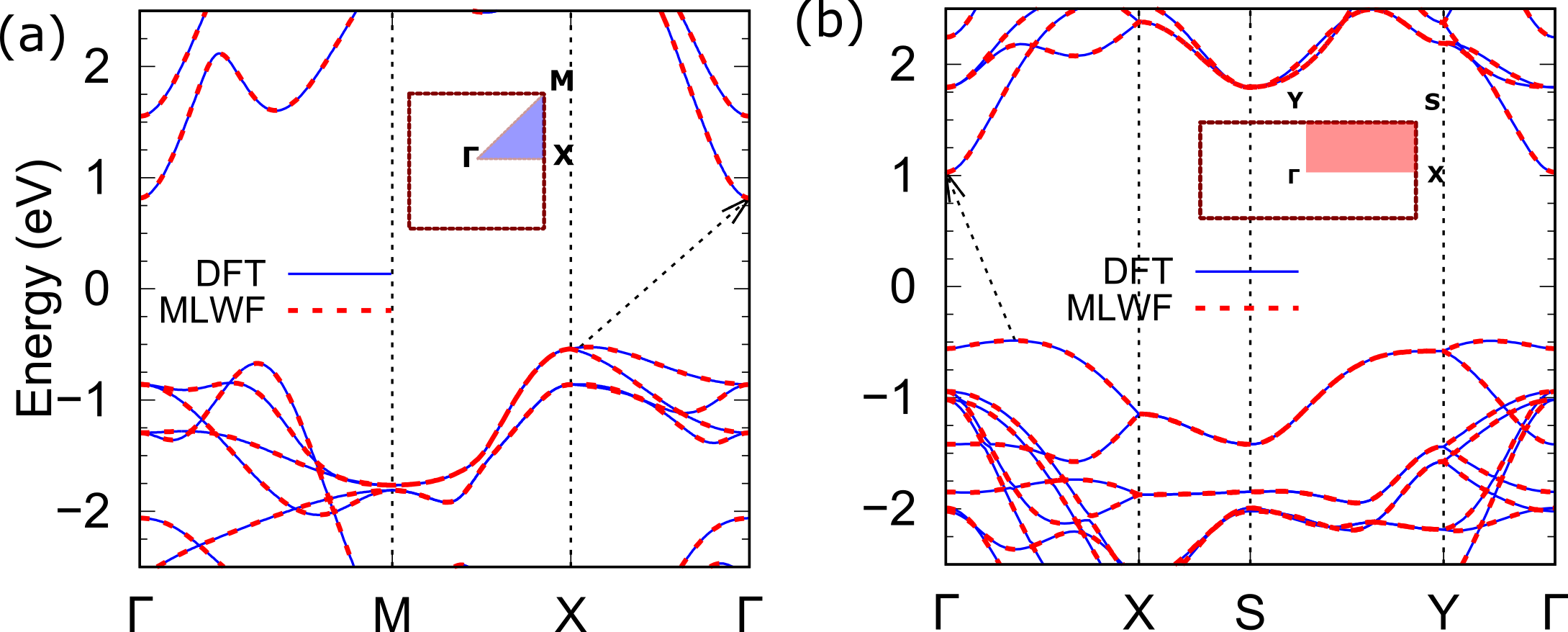}% Here is how to import EPS art
	\caption{\label{BS} Comparison of band structure calculated from DFT and tight-binding
		Hamiltonian (in MLWFs basis) along high symmetry paths in the Brillouin zone for
		(a) Bi$_2$O$_2$Se and (b) InSe. The band structure from MLWFs Hamiltonian shows good
		agreement with DFT in the vicinity of conduction band minima and valance band maxima. Fermi energy level is shifted to zero energy.}
\end{figure}

	\begin{itemize}
	\item The fringing field from the nanowire gate enhances gate efficiency, which in turn improves gate control over the channel for a given gate oxide. When the spacer material is changed from air to SiO$_2$, gate efficiency improves by 75\%, leading to better switching characteristics for the devices.
	
	\item Slight performance improvements are observed when using CaF$_2$ as both the spacer and gate oxide, compared to SiO$_2$ (as both spacer and gate oxide) with same EOT. This is mainly due to the increased physical thickness of the CaF$_2$ gate oxide, which reduces the benefits of having a higher-$k$ spacer (see Fig. \ref{gate_stack} and Table \ref{Table1}). Although a higher-$k$ spacer (with greater permittivity than CaF$_2$) can provide some advantages, it also introduces increased cross-talk, which is undesirable. Moreover, reducing the gate oxide thickness proves beneficial for a given spacer.
	
	\item As the gate oxide thickness increases, the gate's control over the channel weakens due to reduced gate oxide capacitance and decreased fringing capacitance from the gate. This results in a logarithmic relationship between the subthreshold slope and gate oxide thickness, contrasting with conventional FETs \cite{8451970, yoon2011good}.
	\item Increasing the dimensions of the gate uniformly improves switching metrics for a given channel length, and the reverse is also true. However, altering the gate shape from square to triangular degrades the device's performance due to increased direct source-to-drain tunneling. MoSi$_2$N$_4$ demonstrates greater resistance to this type of tunneling. For instance, the subthreshold swing and on-state current deteriorate by approximately 14.5\% and 53\%, respectively, when changing the nanowire gate structure from square to triangular. In contrast, Bi$_2$O$_2$Se and InSe devices experience significant leakage due to severe direct tunneling, attributed to their low carrier effective mass.
	
	\item Ultra-scaled FETs with these promising 2-D semiconducting channels achieve $I_\mathrm{ON}/I_\mathrm{OFF} > 10^3$ with $SS$ $<$ 130 mV/decade. Bi$_2$O$_2$Se stands out, closely followed by InSe and MoSi$_2$N$_4$.
\end{itemize}

\section{Methodology}
To model the full-band transport in the devices, an accurate model Hamiltonian for the channel is required \cite{9585026}. For this, the electronic properties of channel material are characterized from first-principle by combining the power of density-functional theory (DFT) \cite{PhysRevB.54.11169} and maximally localized Wannier functions (MLWFs) \cite{W90}. 
%The modeled system can be up-scaled to match the targeted device dimensions. 
The optimized structural parameters and bandgap values agree well with literature. All these are indirect bandgap semiconductors, with bandgap value lies in range 1.3-1.9 eV; InSe $\sim$ 1.52 eV, Bi$_2$O$_2$Se $\sim$ 1.34 eV, MoSi$_2$N$_4$ $\sim$ 1.84 eV.
Model Hamiltonian validation against DFT results involves calculating eigenenergies along high symmetry points. The band structure from our model demonstrates good agreement with DFT, particularly around the bandgap (for example, see Fig. \ref{BS} for Bi$_2$O$_2$Se and InSe).

\begin{figure}[!t]
	\centering
\includegraphics[width = 0.45\textwidth]{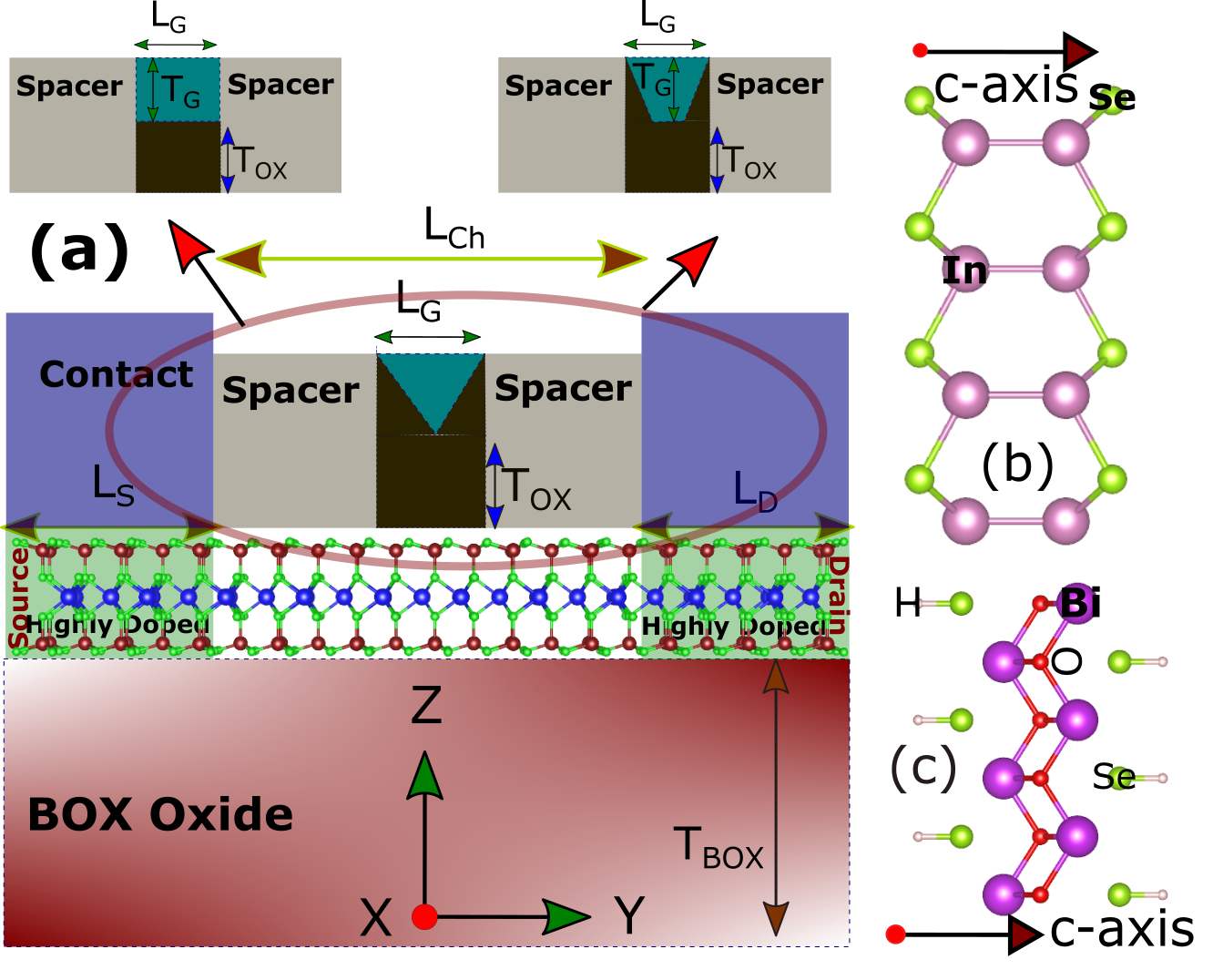}
	\caption{\label{fig2}(a) Schematic of top-gated monolayer 2-D FETs with different gate structures like square, rectangle, isosceles, and triangle. The channel comprises of monolayer MoSi$_2$N$_4$. SiO$_2$ is used as BOX oxide with a thickness of $T_\mathrm{BOX}$. The gate is of metal and separated from the channel by $T_\mathrm{OX}$. (b) Top and side view of atomic structure of MoSi$_2$N$_4$, Bi$_2$O$_2$Se, and InSe.
	}
\end{figure}

The charge density in the channel is obtained by estimating the Green's functions in the non-equilibrium Green's function (NEGF) formalism \cite{datta2005quantum}. The channel charge is included in Poisson's equation to determine the converged physical observables of the system \cite{NanoTCAD}. Green's function is obtained by uniformly sampling the transverse wavevectors with 30 sampling points. The source-to-drain current is calculated by Landauer-B$\mathrm{\ddot{u}}$ttiker approach \cite{Landauer}. The device structure is shown in Fig. \ref{fig2} (a), which is analogous to experimental structures in \cite{Desai99, xiao2022locally}. Considering the ultra-low contact resistance and stability of 2-D device contacted with semimetal, we mimic the degenerate state for the semiconductor in contact with semimetal by chemically doping the source and drain region in our simulations \cite{jiang2023ballistic, shen2021ultralow, li2023approaching, wu2022multiscale}. These results should be regarded as the upper limit for device performance, as non-ideal effects are not considered here. All device simulations are carried out at room temperature.

\section{Results and Discussion}
To capture the switching behavior in these ultra-scaled logic devices, we begin with the monolayer MoSi$_2$N$_4$ channel. The gate dimension are 1 nm $\times$ 1 nm and equivalent oxide thickness (EOT) is 1 nm. The transfer characteristics on semilogarithmic scale are shown in Fig. \ref{FIG3} (a) for air spacer with SiO$_2$ gate dielectric, SiO$_2$ spacer with SiO$_2$ gate dielectric, and calcium fluoride (CaF$_2$) spacer with CaF$_2$ gate dielectric with $L_\mathrm{Ch}$ = 12 nm.
Ensuring a high-quality integration of 2-D channels with high-$k$ oxides is paramount. Epitaxial CaF$_2$ can establish a quasi-van der Waals contact with 2-D semiconductors \cite{illarionov2020insulators, illarionov2019reliability}, resulting in excellent interface quality. Also, it can serve as an ultrathin gate insulator with a high dielectric constant ($\epsilon$ = 8.43) for 2-D devices with an equivalent oxide thickness (EOT) down to 1 nm and minimal leakage current ($I_\mathrm{ON}$/$I_\mathrm{OFF}$ $>$ $10^7$) \cite{illarionov2019ultrathin}. For EOT values of 0.5 nm and 1 nm, 2-D devices with CaF$_2$ gate stack show lower leakage than other high-$k$ oxides, like HfO$_2$ and La$_2$O$_3$ \cite{illarionov2019ultrathin}.
Conversely, defects can adversely impact metal-oxide-semiconductor field-effect transistors (MOSFETs) by causing issues such as charge trapping and de-trapping, bias temperature instabilities (BTI) (driven by charge de-trapping), and mobility degradation (due to coulomb scattering from trapped charges in the oxide) \cite{illarionov2020insulators}.\\

\begin{figure}[!t]
	\centering
	\includegraphics[width = 0.45\textwidth]{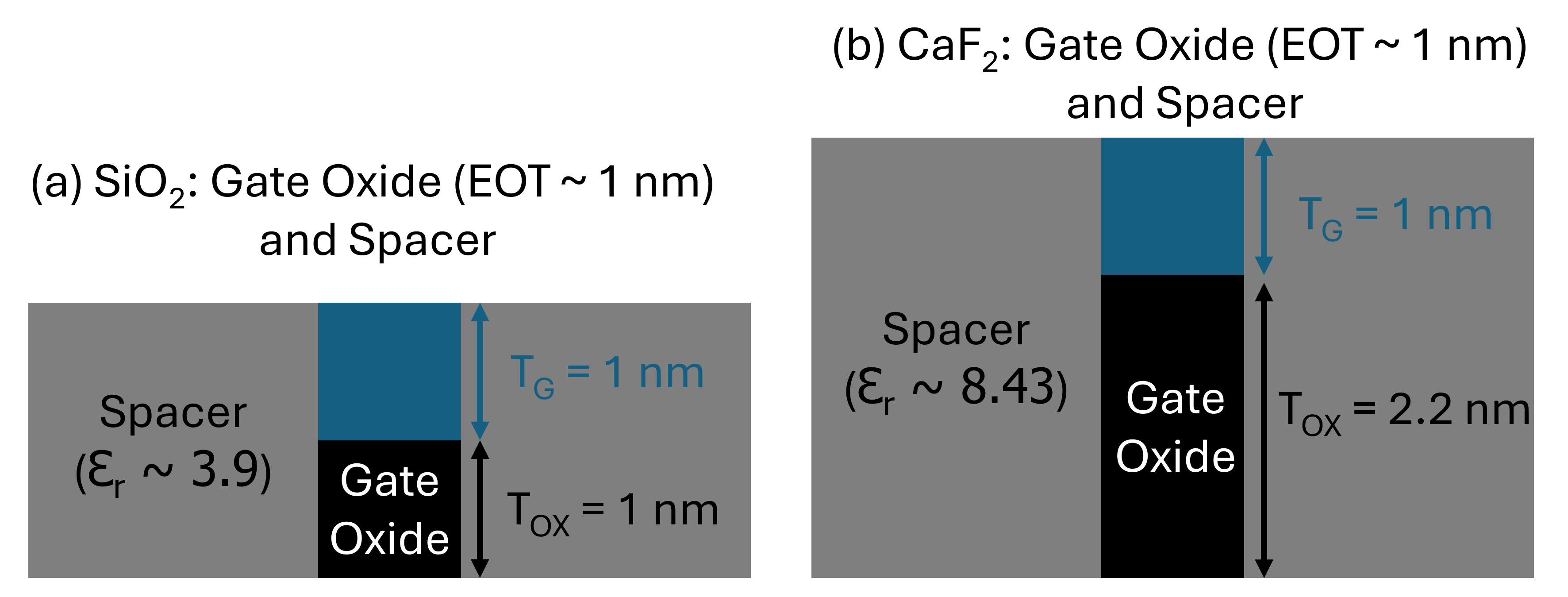}% Here is how to import EPS art
	\caption{\label{gate_stack} The relative comparison between (a) the SiO$_2$ spacer with a SiO$_2$ gate dielectric and (b) the CaF$_2$ spacer with a CaF$2$ gate dielectric, both having the same EOT of 1 nm, shows that the increased $T_{OX}$ in (b) results in only a slight improvement in gate control compared to (a).
		}
\end{figure}

\subsection{Role of Gate Oxide and Spacer}
\begin{figure}[!t]
	\centering
	\includegraphics[width = 0.5\textwidth]{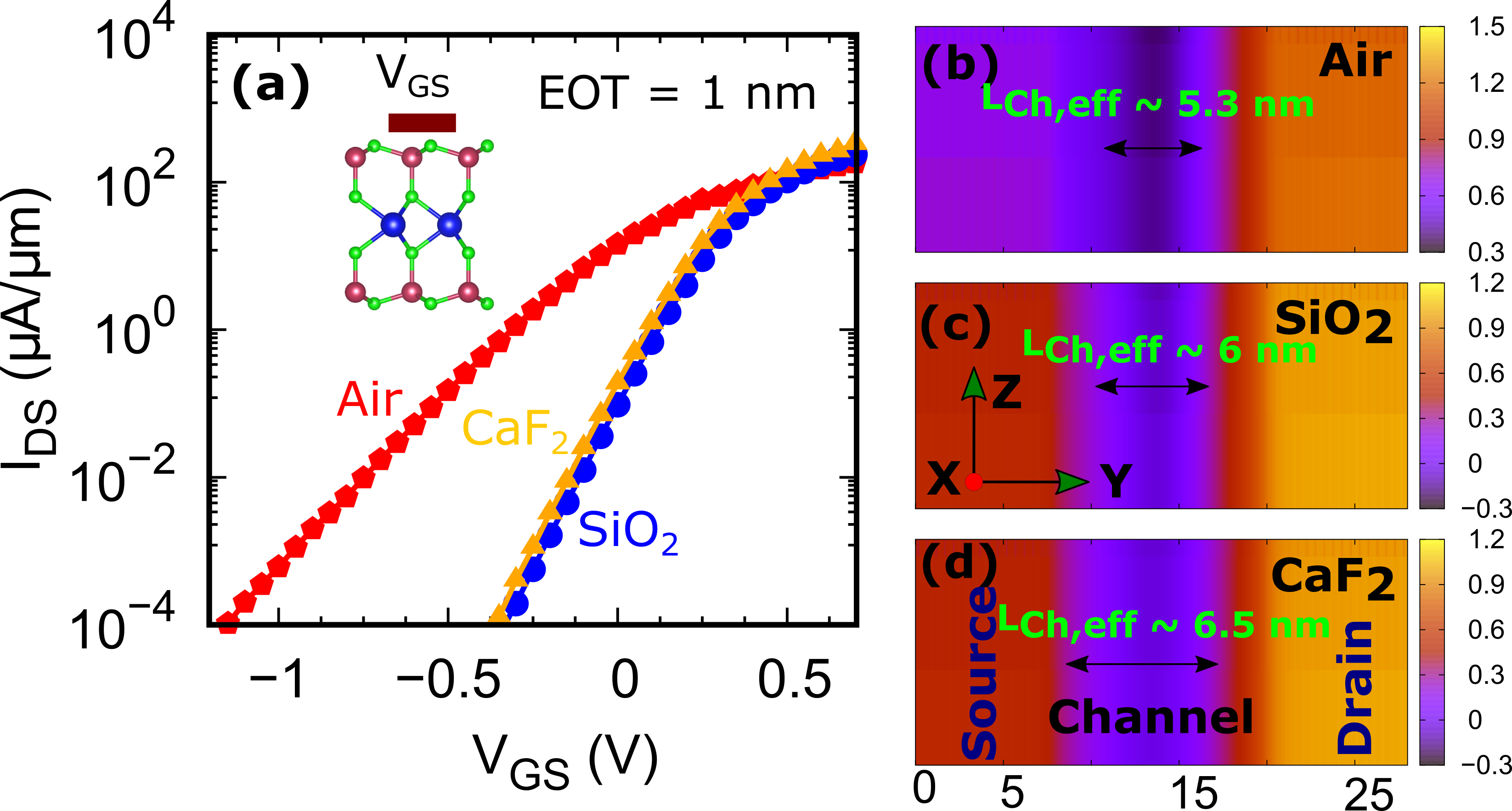}
	\caption{\label{FIG3}(a) Transfer characteristics of ultra-short 2-D FETs comprised of MoSi$_2$N$_4$ for air spacer with SiO$_2$ gate dielectric, SiO$_2$ spacer with SiO$_2$ gate dielectric, and CaF$_2$ spacer with CaF$_2$ gate dielectric. $L_\mathrm{Ch}$ $\sim$ 12 nm and $T_\mathrm{G} \times L_\mathrm{G}$ = 1 nm $\times$ 1 nm. The surface plot of the potential profile in the source, channel, and drain for (b) air spacer, (c) SiO$_2$ spacer, and (d) CaF$_2$ spacer at $V_\mathrm{DS}$ = 0.5 V and $V_\mathrm{GS}$ = -0.5 V. An increase in the dielectric constant of the spacer results in better gate efficiency and, therefore, a larger effective channel length, $L_\mathrm{Ch, eff}$. The device with an air spacer has the worst gate efficiency.
	}
\end{figure}

We observe that effective channel length ($L_\mathrm{eff}$) increases (see Fig. \ref{FIG3} (b-d)) and switching characteristics improves with increase in spacer's permittivity. This improvement is attributed to increased parasitic capacitance arising from the fringing field from gate \cite{5191116}, $C_\mathrm{par}$ $\sim$ $\epsilon_\mathrm{spacer} \times \ln (1 + T_\mathrm{G} / T_\mathrm{ox})$, where $\epsilon_{spacer}$ is the permittivity of spacer. 
The device with air spacer shows poor switching characteristics with gate efficiency $< 0.3$ and subthreshold slope, $SS$ $\sim$ 236 mV/decade with the ON to the OFF current ratio,
 $I_\mathrm{ON}$/$I_\mathrm{OFF}$ $>$ $10^2$. 
 Gate efficiency is defined as $-\Delta \phi_b /\Delta V_{GS}$ in subthreshold regime, where $\phi_b$ is source-to-channel barrier height. 
The simulations are carried out at $V_\mathrm{DS}$ = 0.5 V, where $V_\mathrm{DS}$ denotes the drain to source voltages. The $V_\mathrm{OFF}$ voltage is the gate voltage, $V_\mathrm{GS}$, that produces a drain current $I_\mathrm{DS}$ equal to a specified $I_\mathrm{OFF}$ $\sim$ $10^{-1}$ $\mu$A/$\mu$m. The $I_\mathrm{ON}$ of a FET is measured at the ON-state voltage $V_\mathrm{ON}=V_\mathrm{OFF}+V_\mathrm{DD}$, where $V_\mathrm{DD}$ = 0.5 V is the power supply voltage.
Replacing the spacer with higher permittivity improves the switching characteristics. For example, SiO$_2$ spacer improves the gating efficiency by 75\%, therefore better-switching characteristics ($SS$ $\sim$ 129 mV/decade) and $I_\mathrm{ON}$/$I_\mathrm{OFF}$ ($>$ $10^3$). However, small performance improvements are observed in the case of CaF$_2$ as both spacer and gate oxide, compared to SiO$_2$ (as both spacer and gate oxide). This is primarily due
	to its increased gate oxide physical thickness, which diminish the impact of having higher-$k$ spacer (see Fig. \ref{gate_stack} and Table. \ref{Table1}). Building on this analysis, we further demonstrate the impact of spacer with CaF$_2$ gate
	oxide. The results indicate that while a higher-
	$k$ spacer (with greater permittivity than CaF$_2$) can
	offer advantages, it comes at the cost of increased cross-talk, which is undesirable. Additionally, reducing the gate oxide thickness proves advantageous within the given spacer configuration.

\begin{table}[!b]
	\caption{Subthreshold slope ($SS$), ON-state current ($I_\mathrm{ON}$), and ON current to the OFF current ratio ($I_\mathrm{ON}$/$I_\mathrm{OFF}$) for different combinations of the gate oxides and Spacers.} \label{Table1}
\centering
%	\resizebox{\textwidth}{!}
\begin{tabular}{c c c c c c c c}
	\hline \\[0.000005ex]
	Gate Oxide & Spacer & $SS$ & $I_\mathrm{ON}$  & $I_\mathrm{ON}$/$I_\mathrm{OFF}$ \\
	
	&         & (mV/decade) & ($\mu A/ \mu m$) & \\[1ex]   
	\hline \\[0.0005ex]
	SiO$_2$   & Air & 236 & 18.45 & 1.84 $\times 10^2$  \\[1ex]
	
	SiO$_2$   & SiO$_2$  & 128.98 & 106.67 & 1.07 $\times 10^{3}$\\[1ex]
	
	CaF$_2$ &  CaF$_2$  & 124.90 & 117.62 & 1.18 $\times$ $10^3$ \\[1ex]

	\hline 
	
\end{tabular}
\end{table}

\begin{figure}[!t]
	\centering
	\includegraphics[width = 0.4\textwidth]{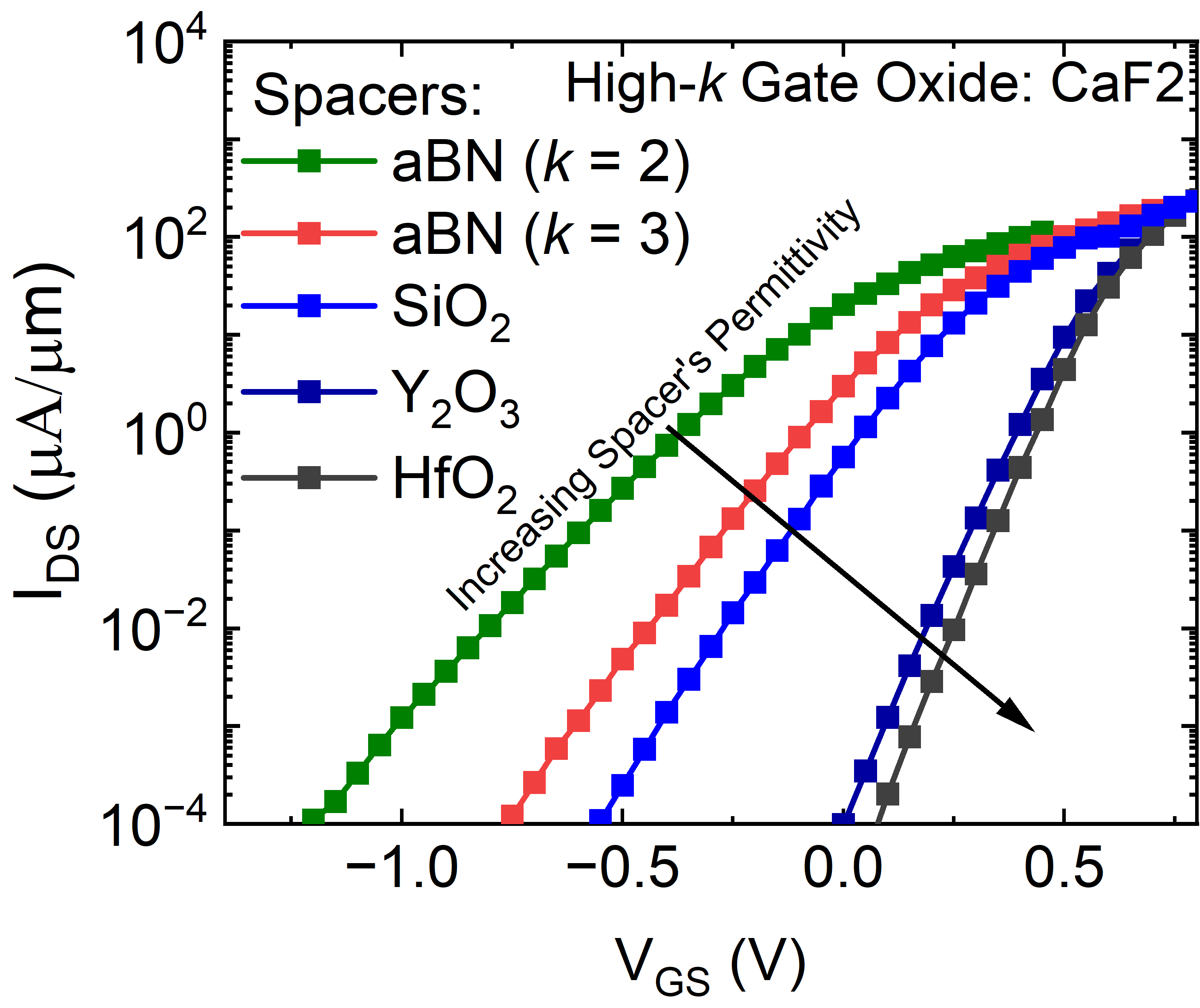}% Here is how to import EPS art
	\caption{\label{Diff_spacer} Transfer characteristics of ultra-short 2-D FETs, showing the role of spacer.
		The channel is comprised of MoSi$_2$N$_4$, CaF$_2$ is high-$k$ gate dielectric. $L_\mathrm{Ch}$ $\sim$ 12 nm and $T_\mathrm{G} \times L_\mathrm{G}$ = 1 nm $\times$ 1 nm. Small performance gains are observed when $k$ $>$ 8, though this range is less favorable due to increased electronic cross-talk.}
\end{figure}

\begin{figure}[!t]
	\centering
	\includegraphics[width = 0.4\textwidth]{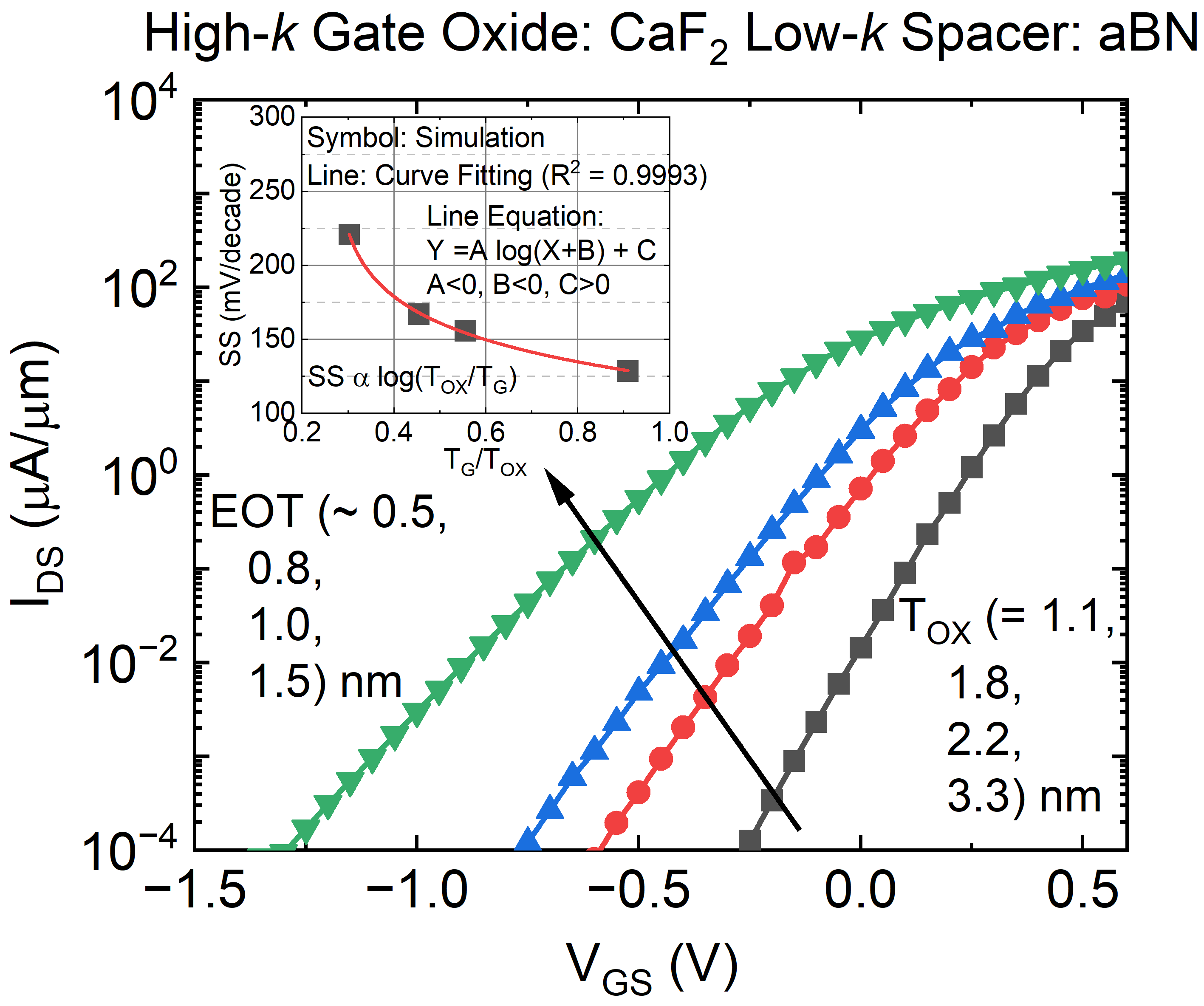}% Here is how to import EPS art
	\caption{\label{Diff_TOX} Transfer characteristics of ultra-short 2-D FETs, showing the role gate oxide thickness scaling.
		The channel is comprised of MoSi$_2$N$_4$, CaF$_2$ is high-$k$ gate dielectric, and amorphous boron nitride (aBN) is low-$k$ spacer. $L_\mathrm{Ch}$ $\sim$ 12 nm and $T_\mathrm{G} \times L_\mathrm{G}$ = 1 nm $\times$ 1 nm. In the inset, $SS$ is plotted with $T_G$/$T_{OX}$. Unlike conventional FETs, this trend is logarithmic with $T_{OX}$.}
\end{figure}

Further, we assessed the performance of ultra-short devices with various spacers, ranging from low-$k$ to high-$k$ materials, using CaF$_2$ as a high-k gate dielectric (see Fig. \ref{Diff_spacer}), which is ideally suited for 2-D channel materials. Performance improvements are noted when $k$ $>$8, but this range is less favorable due to heightened electronic cross-talk.
To analyze the impact of physical gate oxide thickness, we consider low-$k$ spacer with permittivity of 3 and CaF$_2$ as high-$k$ gate oxide. We perform the analysis with different gate oxide thickness ($T_{OX}$), ranging from 3.3 nm (EOT $\sim$ 1.5 nm) down to 1.1 nm (EOT $\sim$ 0.5 nm). As the oxide thickness ($T_{OX}$) increases for a given gate oxide, gate dimensions, and spacer configuration, the gate's control over the channel weakens. This occurs due to a combination of effects from the gate oxide capacitance and fringing capacitance caused by the spacer. The gate oxide capacitance decreases when the gate oxide thickness increases, reducing the gate's control. Additionally, the fringing capacitance from the spacer plays a role in gate control in short-channel devices. An increase in $T_{OX}$ decreases parasitic capacitance, further weakening the gate's control, and conversely, reducing $T_{OX}$ increases parasitic capacitance, enhancing the gate's control over the channel.
To show the impact of $T_{OX}$ quantitatively, we plot the transfer characteristics with different $T_{OX}$ ranging from 3.3 nm down to 1.1 nm (see Fig. \ref{Diff_TOX}). This $T_{OX}$ variation is equivalent to EOT's variation from 1.5 nm down to 0.5 nm. The channel is comprised of MoSi$_2$N$_4$, CaF$_2$ is used as gate dielectric, and aBN is used as low-$k$ spacer. The $SS$ is plotted with $T_G$/$T_{OX}$ in inset of Fig. \ref{Diff_TOX}. Unlike conventional FETs \cite{8451970, yoon2011good}, the $SS$ vs $T_{OX}$ trend is logaritmic with $T_{OX}$ due to the role of fringing in determining the gate control. Further study considers SiO$_2$ to be both a spacer and gate oxide. The gate oxides and spacers are considered perfect insulators.

\subsection{Role of Channel Materials}
\begin{figure}[!t]
	\centering
	\includegraphics[width = 0.49\textwidth]{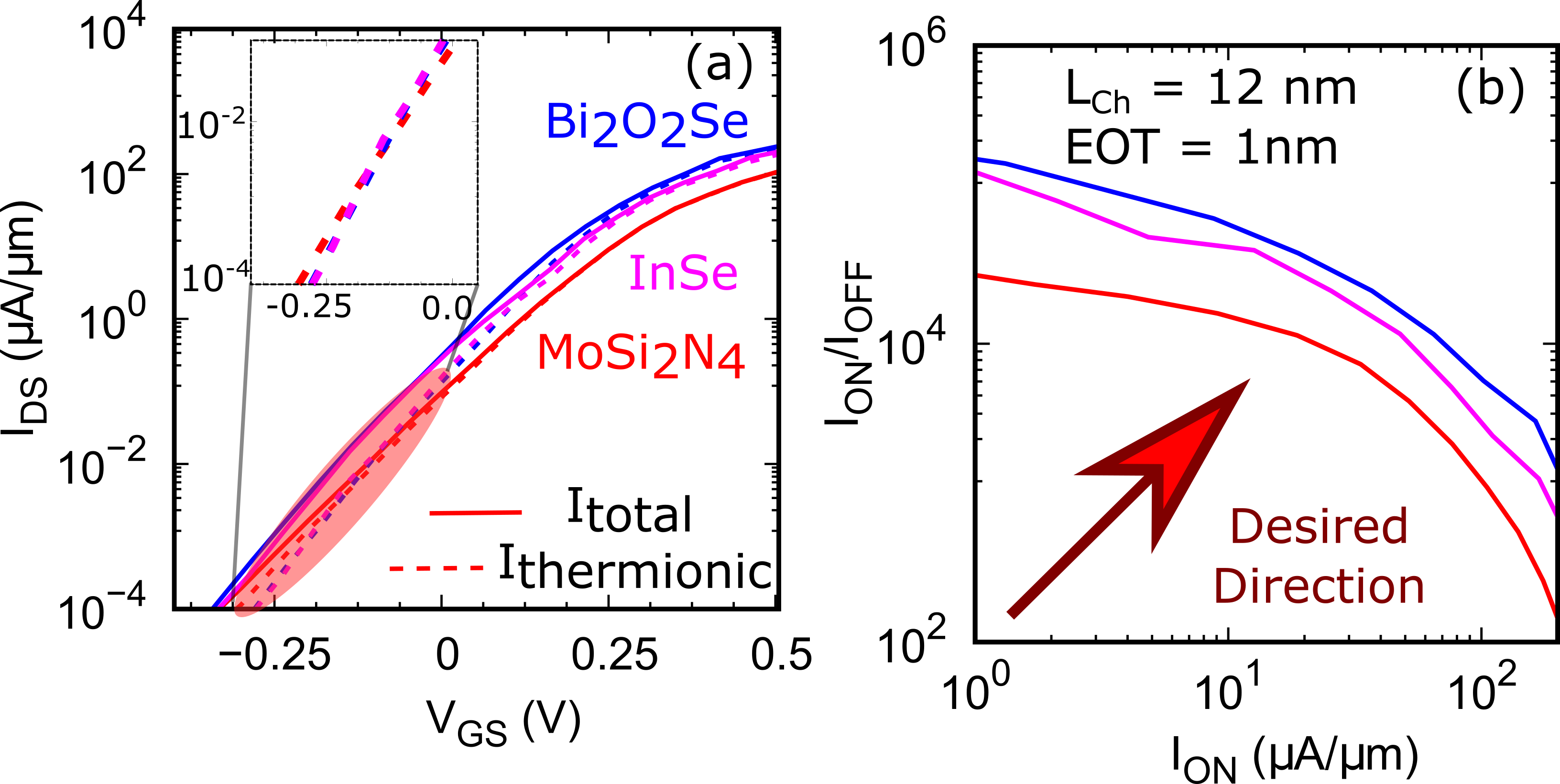}% Here is how to import EPS art
	\caption{\label{Id_Vg_bechmarked}(a) Benchmarking of transfer characteristics and (b) $I_\mathrm{ON}$ vs. $I_\mathrm{ON}/I_\mathrm{OFF}$ for
		three promising 2-D semiconductors as a channel in ultra-short FET architecture. SiO$_2$
		is used as gate oxide with EOT = 1 nm, spacer, and BOX oxide with $T_\mathrm{BOX}$ = 10 nm. 
		The channel length is of 12 nm, $T_\mathrm{G} \times L_\mathrm{G}$ = 1 nm $\times$ 1 nm, and $V_\mathrm{DS}$ = 0.50 V. Bi$_2$O$_2$Se device show best switching characteristic and $I_\mathrm{ON}$ for a given $I_\mathrm{ON}/I_\mathrm{OFF}$.% \vspace{-4mm}
	}
\end{figure}

For 2-D semiconductors, the centroid capacitance is often assumed to be infinite \cite{ma2015carrier}, therefore $C_S$ $\sim$ $C_Q$, where $C_S$ is the capacitance of the channel material and $C_Q$ is the quantum capacitance of the channel material. Therefore, $SS$ is proportional to $(1 + C_\mathrm{Q}/C_\mathrm{OX})$ in the negligible tunneling regime \cite{1224486}. Thus, a lower quantum capacitance channel would be beneficial in reducing the thermionic $SS$ \cite{9286895}.
This is evident from the subthreshold slope of the thermionic current, $SS_\mathrm{thermal}$. It is the best for the Bi$_2$O$_2$Se channel ($\sim 98$ mV/decade) and slightly degrades for InSe ($\sim 100$ mV/decade) compared to Bi$_2$O$_2$Se. As the quantum capacitance values are approximately similar for InSe ($m_\mathrm{e} \sim 0.205m_0$ and $g_v = 1$) and Bi$_2$O$_2$Se ($m_\mathrm{e} \sim 0.197 m_0$ and $g_v = 1$), both being less than one fourth of MoSi$_2$N$_4$ ($m_\mathrm{e} \sim 0.48 m_0$ and $g_v = 2$). For MoSi$_2$N$_4$, the subthreshold slope is the worst, close to 125 mV/decade. The thermionic current component is plotted in Fig. \ref{Id_Vg_bechmarked} (a), providing a zoomed view in the subthreshold regime.
However, at a channel length of $L_\mathrm{Ch} = 12$ nm, the switching behavior is influenced by both direct tunneling leakage and electrostatics. The contribution of direct tunneling from source-to-drain in the off-state current, $I_\mathrm{OFF}$, is significant for InSe ($\sim 54\%$) and Bi$_2$O$_2$Se ($\sim 59\%$) but lower for MoSi$_2$N$_4$ ($\sim 12\%$). The strength of InSe and Bi$_2$O$_2$Se are better electrostatic control in the devices comprised of them. Even in the presence of large direct tunneling, InSe and Bi$_2$O$_2$Se demonstrate better control over the switching behavior compared to MoSi$_2$N$_4$ for $L_\mathrm{Ch}$ = 12 nm (see Fig. \ref{Id_Vg_bechmarked}).

Devices comprises of these promising 2-D semiconducting channel can deliver $I_\mathrm{ON}/I_\mathrm{OFF}$ $>$ $10^3$ with $SS$ $<$ 130 mV/decade. Bi$_2$O$_2$Se is the best-performing channel material (SS $\sim$ 109 mV/decade and $I_\mathrm{ON}$ $\sim$ 186 $\mu$A/$\mu$m) and the performance of the InSe (SS $\sim$ 112 mV/decade and $I_\mathrm{ON}$ $\sim$ 153 $\mu$A/$\mu$m) is close to Bi$_2$O$_2$Se, followed by MoSi$_2$N$_4$. The $I_\mathrm{ON}/I_\mathrm{OFF}$ of all these three channel materials are one order higher than 1 nm MoS$_2$ FET with a circular metallic gate of diameter 1 nm with 10 nm channel length (SS $\sim$ 160 mV/decade and $I_\mathrm{ON}/I_\mathrm{OFF}$ $\sim$ 400) \cite{doi:10.1063/1.5054281}. 
Increasing the $L_\mathrm{Ch}$, degrades the control of gate over channel and on contrary reduce the impact of direct tunneling (because of increased tunneling barrier height and width) and vice versa (see Fig. \ref{Figure_5_2} (a)). This device configuration comprises these novel 2-D semiconductors and can deliver better switching performance than other ultra-short device configurations like FETs with core-shell nanowire gate \cite{cao2016prospects,liu2018mos2}.

In the quest of performance improvement, applying 1\% tensile strain along the channel direction, which is reasonable in experiments with flexible substrates \cite{datye2022strain,song2018largely}, leads to minor change in performance of the FETs compared to pristine channel FETs. As, it can lower the $C_\mathrm{Q}$ by 3.8\%, -0.25\%, and -0.005\% in MoSi$_2$N$_4$, Bi$_2$O$_2$Se, and InSe, respectively, than a pristine channel. The decrease in bandgap is less than 6\% compared to pristine monolayers, which is suitable for mitigating intra-band tunneling leakage. Other possible solution could be utilizing the non-linear dielectric response of dielectric-ferroelectric (DE-FE) \cite{10.1063/5.0056448} stacked or FE-antiferroelectric (FE-AFE) \cite{cheema2022ultrathin} stacked materials in the gate-stack. This is a promising approach to enhance the performance of devices by amplifying the surface potential's response to changes in gate voltage.

\subsection{Role of Gate's dimensions and Performance Metrics}
\begin{figure}[!t]
	\centering
	\includegraphics[width = 0.40\textwidth]{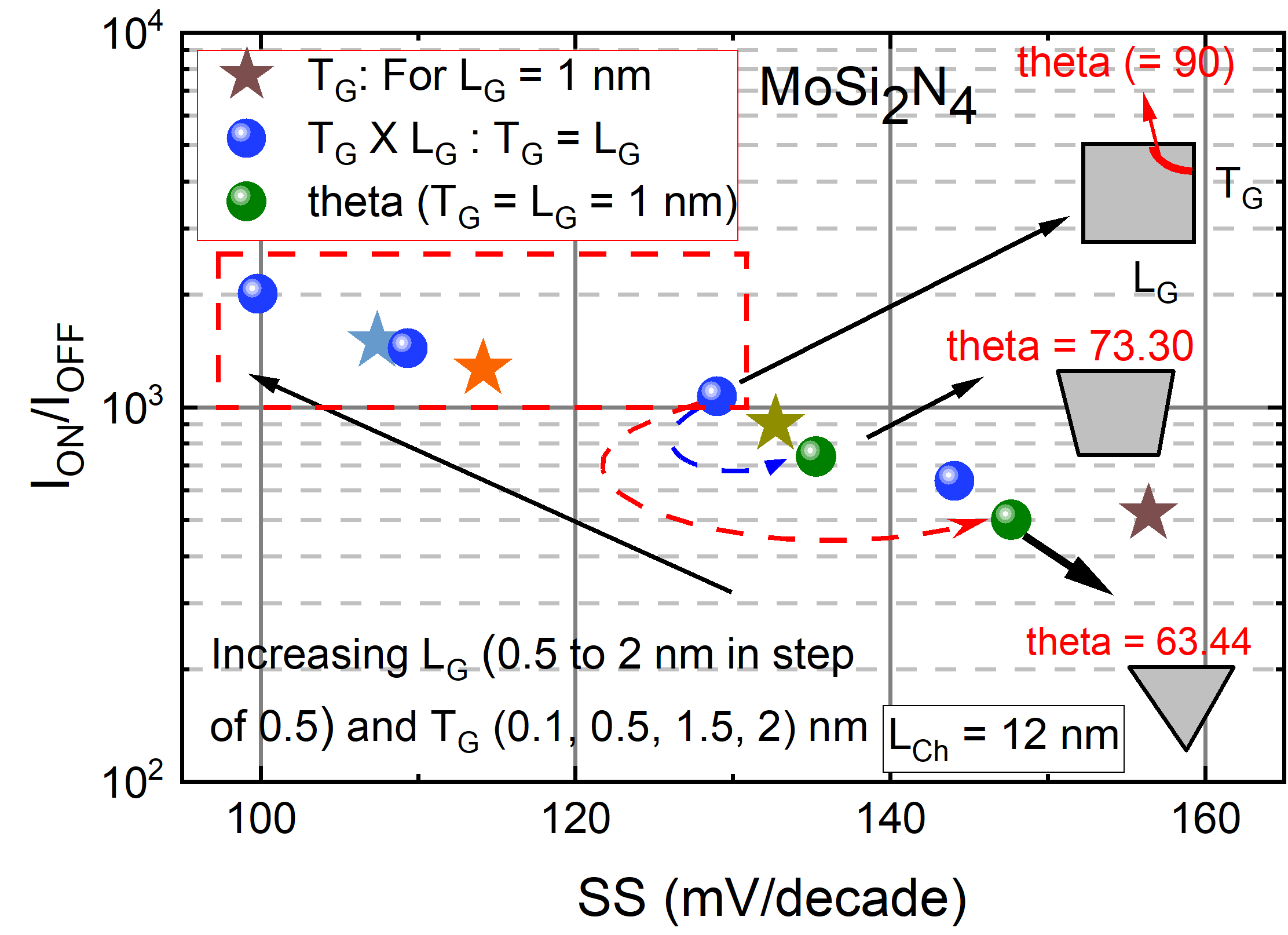}% Here is how to import EPS art
	\caption{\label{Figure_5_1}The role of gate dimensions in the performance of ultra-short channel devices.
	}
\end{figure}

The role of gate geometry and dimensions is also studied for better assessment of device performance over design space. We find that a uniform increment in $T_\mathrm{G}$ and $L_\mathrm{G}$ improves the switching metrics for a given channel length and vice versa.
For example, a change in gate dimensions from 0.5 $\times$ 0.5 to 2 $\times$ 2, improves the gate efficiency by $\sim$ 45\%, $SS$ by $\sim$ 30\%, and $I_\mathrm{ON}$ by $\sim$ 215\% in MoSi$_2$N$_4$ devices. Increasing gate height from 0.1 nm to 2 nm with $L_\mathrm{G}$ of 1 nm enhances the gate efficiency by $\sim$ 34\%, $SS$ by $\sim$ 31\%, and $I_\mathrm{ON}$ by $\sim$ 189\%.
The gate geometry variations, from square to triangular, are characterized by an angle theta (in degrees). For MoSi$_2$N$_4$, the $SS$ and $I_\mathrm{ON}$ are deteriorated by $\sim$ 14.5\% and 53\%, respectively, while varying the gate structure from square (theta = 90) to triangular (theta = 63.44). For theta $>$ 72 (78), the degradation in $I_\mathrm{ON}$ is less than 32\% (23\%), see Fig. \ref{Figure_5_1}.
For Bi$_2$O$_2$Se and InSe devices, severe direct tunneling results in severe leakage for theta $<$ 73.30 due to low carrier effective mass.

\begin{figure}[!t]
	\centering
	\includegraphics[width = 0.40\textwidth]{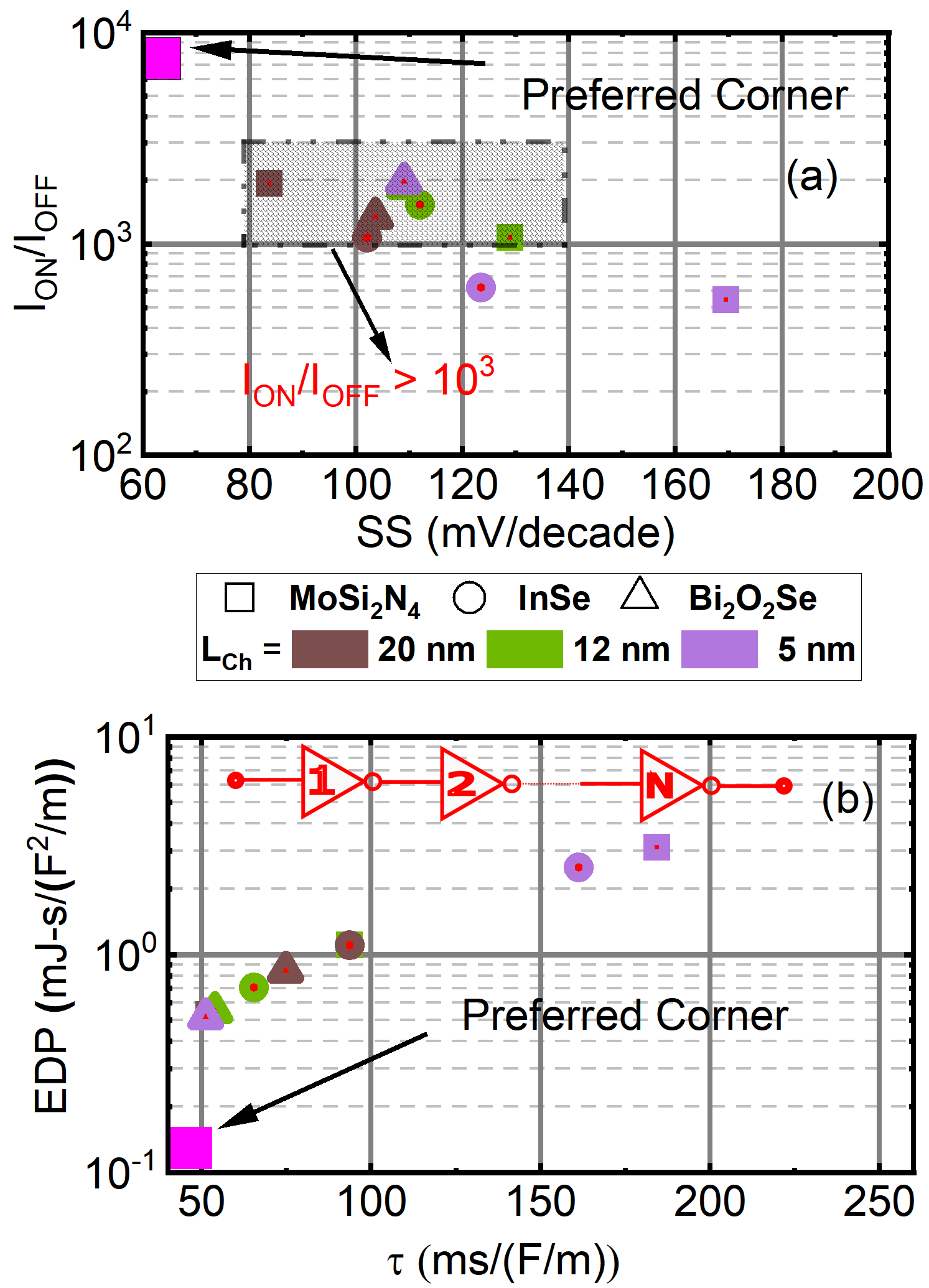}% Here is how to import EPS art
	\caption{\label{Figure_5_2}
		(a) Benchmarking of switching characteristics (SS vs $I_\mathrm{ON}/I_\mathrm{OFF}$) for the devices. (b) Circuit level characteristics ($\tau$ vs EDP) for a 20 stage inverter comprised of these ultra-short 2-D channel devices.
	}
\end{figure}

Finally, critical circuit-level performance metrics are estimated for a chain of inverters with a logic depth of 20, as shown in Fig. \ref{Figure_5_2}. We have estimated the delay and energy per operation. These metrics have been normalized by the node capacitance and the total capacitance of the logic design. This normalization approach is reasonable in sub-10 nm technology, as interconnect capacitance becomes the dominant component of total capacitance \cite{agarwal2016effect, 10091159}.
For these promising channel materials, the normalized delay ($\tau$) and energy-delay product (EDP) for a chain of 20 inverters are impressively low, measuring less than 95 and 1.5, respectively. To put this into perspective, these circuit-level metrics outperform 1 nm MoS$_2$ FET with a circular metallic gate of diameter 1 nm with 10 nm channel length (the $\tau$ is around 250, and the EDP is approximately 5 \cite{doi:10.1063/1.5054281}).

\section{Conclusion}
In the pursuit of identifying promising 2-D semiconductors for cutting-edge electronic applications, we have identified MoSi$_2$N$_4$, InSe, and Bi$_2$O$_2$Se as highly promising candidates.
By considering the feasibility of integrating metallic nanowires with 2-D semiconductors, we have investigated their suitability as channel materials within an innovative device structure. Our goal is to harness their unique properties and capabilities for ultra-scaled FETs with ultra-scaled gate lengths (approximately 1 nm).
We have found that the Bi$_2$O$_2$Se FET exhibits the most favorable switching metrics and circuit-level performance. InSe FETs perform closely to Bi$_2$O$_2$Se, with MoSi$_2$N$_4$ following closely behind.

Beyond its applications in logic circuits, this novel device structure has the potential to be extended into a multi-gate configuration model \cite{xiao2022locally}. This extension opens up exciting possibilities for achieving precise and localized control over band structures, which in turn enables the design of more integrated, multifunctional, and highly controllable nano-devices. Moreover, this versatile device topology can also find utility in the development of ultra-low-power spiking neurons, particularly for neuromorphic applications \cite{thakar2023ultra}. This represents an exciting direction where the technology can contribute to the advancement of brain-inspired computing.

\section{Acknowledgment}
The authors would like to thank HPC facility, IIT Kanpur, Kanpur, India and the National Supercomputing Mission (NSM) for providing the computational resources. KN is thankful to Ashok P. for the insightful discussions.

%\clearpage
\balance
\bibliographystyle{IEEEtranDOI}
\bibliography{asd}

\providecommand{\noopsort}[1]{}\providecommand{\singleletter}[1]{#1}%
\begin{thebibliography}{10}
\providecommand{\url}[1]{#1}
\csname url@samestyle\endcsname
\providecommand{\newblock}{\relax}
\providecommand{\bibinfo}[2]{#2}
\providecommand{\BIBentrySTDinterwordspacing}{\spaceskip=0pt\relax}
\providecommand{\BIBentryALTinterwordstretchfactor}{4}
\providecommand{\BIBentryALTinterwordspacing}{\spaceskip=\fontdimen2\font plus
\BIBentryALTinterwordstretchfactor\fontdimen3\font minus
  \fontdimen4\font\relax}
\providecommand{\BIBforeignlanguage}[2]{{%
\expandafter\ifx\csname l@#1\endcsname\relax
\typeout{** WARNING: IEEEtran.bst: No hyphenation pattern has been}%
\typeout{** loaded for the language `#1'. Using the pattern for}%
\typeout{** the default language instead.}%
\else
\language=\csname l@#1\endcsname
\fi
#2}}
\providecommand{\BIBdecl}{\relax}
\BIBdecl

\bibitem{zhuo2023modifying}
\BIBentryALTinterwordspacing
F.~Zhuo, J.~Wu, B.~Li, M.~Li, C.~L. Tan, Z.~Luo, H.~Sun, Y.~Xu, and Z.~Yu,
  ``Modifying the power and performance of 2-dimensional {MoS}$_2$ field effect
  transistors,'' \emph{Research}, vol.~6, p. 0057,   2023, doi:
  \href{http://dx.doi.org/10.34133/research.0057}{{10.34133/research.0057}}.
\BIBentrySTDinterwordspacing

\bibitem{10.3389/felec.2023.1277927}
\BIBentryALTinterwordspacing
K.~Nandan, A.~Agarwal, S.~Bhowmick, and Y.~S. Chauhan, ``Two-dimensional
  semiconductors based field-effect transistors: review of major milestones and
  challenges,'' \emph{Frontiers in Electronics}, vol.~4,   2023, doi:
  \href{http://dx.doi.org/10.3389/felec.2023.1277927}{{10.3389/felec.2023.1277927}}.
\BIBentrySTDinterwordspacing

\bibitem{cao2016prospects}
W.~Cao, W.~Liu, and K.~Banerjee, ``Prospects of ultra-thin nanowire gated
  2d-fets for next-generation cmos technology,'' in \emph{2016 IEEE
  International Electron Devices Meeting (IEDM)},  2016, doi:
  \href{http://dx.doi.org/10.1109/IEDM.2016.7838419}{{10.1109/IEDM.2016.7838419}}.
  pp. 14.7.1--14.7.4.

\bibitem{Desai99}
\BIBentryALTinterwordspacing
S.~B. Desai, S.~R. Madhvapathy, A.~B. Sachid, J.~P. Llinas, Q.~Wang, G.~H. Ahn,
  G.~Pitner, M.~J. Kim, J.~Bokor, C.~Hu, H.-S.~P. Wong, and A.~Javey,
  ``{MoS}$_2$ transistors with 1-nanometer gate lengths,'' \emph{Science}, vol.
  354, no. 6308, pp. 99--102,   2016, doi:
  \href{http://dx.doi.org/10.1126/science.aah4698}{{10.1126/science.aah4698}}.
\BIBentrySTDinterwordspacing

\bibitem{wu2022vertical}
\BIBentryALTinterwordspacing
F.~Wu, H.~Tian, Y.~Shen, Z.~Hou, J.~Ren, G.~Gou, Y.~Sun, Y.~Yang, and T.-L.
  Ren, ``Vertical {MoS}$_2$ transistors with sub-1-nm gate lengths,''
  \emph{Nature}, vol. 603, no. 7900, pp. 259--264,   Mar 2022, doi:
  \href{http://dx.doi.org/10.1038/s41586-021-04323-3}{{10.1038/s41586-021-04323-3}}.
\BIBentrySTDinterwordspacing

\bibitem{xiao2022locally}
Y.~Xiao, G.~Zou, J.~Huo, T.~Sun, B.~Feng, and L.~Liu, ``Locally thinned,
  core–shell nanowire-integrated multi-gate {MoS}$_2$ transistors for active
  control of extendable logic,'' \emph{ACS Applied Materials \& Interfaces},
  vol.~15, no.~1, pp. 1563--1573,  2023, doi:
  \href{http://dx.doi.org/10.1021/acsami.2c17788}{{10.1021/acsami.2c17788}}.

\bibitem{liu2018mos2}
X.~Liu, R.~Liang, G.~Gao, C.~Pan, C.~Jiang, Q.~Xu, J.~Luo, X.~Zou, Z.~Yang,
  L.~Liao, and Z.~L. Wang, ``{MoS}$_2$ negative-capacitance field-effect
  transistors with subthreshold swing below the physics limit,'' \emph{Advanced
  Materials}, vol.~30, no.~28, p. 1800932,  2018, doi:
  \href{http://dx.doi.org/10.1002/adma.201800932}{{10.1002/adma.201800932}}.

\bibitem{jolie20241d}
W.~Jolie and T.~Michely, ``1d metals for 2d electronics,'' \emph{Nature
  Nanotechnology}, pp. 1--2, 2024.

\bibitem{Hong}
Y.-L. Hong, Z.~Liu, L.~Wang, T.~Zhou, W.~Ma, C.~Xu, S.~Feng, L.~Chen, M.-L.
  Chen, D.-M. Sun, X.-Q. Chen, H.-M. Cheng, and W.~Ren, ``Chemical vapor
  deposition of layered two-dimensional {MoSi$_2$N$_4$} materials,''
  \emph{Science}, vol. 369, no. 6504, pp. 670--674,  2020, doi:
  \href{http://dx.doi.org/10.1126/science.abb7023}{{10.1126/science.abb7023}}.

\bibitem{Wang2021}
Q.~Wang, L.~Cao, S.-J. Liang, W.~Wu, G.~Wang, C.~H. Lee, W.~L. Ong, H.~Y. Yang,
  L.~K. Ang, S.~A. Yang, and Y.~S. Ang, ``Efficient ohmic contacts and built-in
  atomic sublayer protection in mosi$_2$n$_4$ and wsi$_2$n$_4$ monolayers,''
  \emph{npj 2D Materials and Applications}, vol.~5, no.~1, p.~71,  Aug 2021,
  doi:
  \href{http://dx.doi.org/10.1038/s41699-021-00251-y}{{10.1038/s41699-021-00251-y}}.

\bibitem{9646230}
K.~Nandan, B.~Ghosh, A.~Agarwal, S.~Bhowmick, and Y.~S. Chauhan,
  ``Two-dimensional mosi2n4: An excellent 2-d semiconductor for field-effect
  transistors,'' \emph{IEEE Transactions on Electron Devices}, vol.~69, no.~1,
  pp. 406--413,  2022, doi:
  \href{http://dx.doi.org/10.1109/TED.2021.3130834}{{10.1109/TED.2021.3130834}}.

\bibitem{PhysRevApplied.19.064058}
\BIBentryALTinterwordspacing
K.~Nandan, S.~Bhowmick, Y.~S. Chauhan, and A.~Agarwal, ``Designing
  power-efficient transistors using narrow-bandwidth materials from the
  $m{A}_{2}{Z}_{4}$ ($m=\text{Mo, Cr, Zr, Ti, Hf}$; $a=\text{Si, Ge}$;
  $z=\text{N, P, As}$) monolayer series,'' \emph{Phys. Rev. Appl.}, vol.~19, p.
  064058,   Jun 2023, doi:
  \href{http://dx.doi.org/10.1103/PhysRevApplied.19.064058}{{10.1103/PhysRevApplied.19.064058}}.
\BIBentrySTDinterwordspacing

\bibitem{li20212d}
T.~Li and H.~Peng, ``{2D} {Bi$_2$O$_2$Se}: An emerging material platform for
  the next-generation electronic industry,'' \emph{Accounts of Materials
  Research}, vol.~2, no.~9, pp. 842--853,  2021, doi:
  \href{http://dx.doi.org/10.1021/accountsmr.1c00130}{{10.1021/accountsmr.1c00130}}.

\bibitem{quhe2019high}
R.~Quhe, J.~Liu, J.~Wu, J.~Yang, Y.~Wang, Q.~Li, T.~Li, Y.~Guo, J.~Yang,
  H.~Peng \emph{et~al.}, ``High-performance sub-10 nm monolayer bi 2 o 2 se
  transistors,'' \emph{Nanoscale}, vol.~11, no.~2, pp. 532--540, 2019.

\bibitem{zhang2022single}
Y.~Zhang \emph{et~al.}, ``A single-crystalline native dielectric for
  two-dimensional semiconductors with an equivalent oxide thickness below
  0.5{\thinspace}nm,'' \emph{Nature Electronics}, vol.~5, no.~10, pp. 643--649,
   Oct 2022, doi:
  \href{http://dx.doi.org/10.1038/s41928-022-00824-9}{{10.1038/s41928-022-00824-9}}.

\bibitem{tan20232d}
C.~Tan, M.~Yu, J.~Tang, X.~Gao, Y.~Yin, Y.~Zhang, J.~Wang, X.~Gao, C.~Zhang,
  X.~Zhou, L.~Zheng, H.~Liu, K.~Jiang, F.~Ding, and H.~Peng, ``{2D} fin
  field-effect transistors integrated with epitaxial high-k gate oxide,''
  \emph{Nature}, vol. 616, no. 7955, pp. 66--72,  Apr 2023, doi:
  \href{http://dx.doi.org/10.1038/s41586-023-05797-z}{{10.1038/s41586-023-05797-z}}.

\bibitem{jiang2023ballistic}
J.~Jiang, L.~Xu, C.~Qiu, and L.-M. Peng, ``Ballistic two-dimensional {InSe}
  transistors,'' \emph{Nature}, vol. 616, no. 7957, pp. 470--475,  Apr 2023,
  doi:
  \href{http://dx.doi.org/10.1038/s41586-023-05819-w}{{10.1038/s41586-023-05819-w}}.

\bibitem{song2023wafer}
S.~Song, S.~Jeon, M.~Rahaman, J.~Lynch, D.~Rhee, P.~Kumar, S.~Chakravarthi,
  G.~Kim, X.~Du, E.~W. Blanton, K.~Kisslinger, M.~Snure, N.~R. Glavin, E.~A.
  Stach, R.~H. Olsson, and D.~Jariwala, ``Wafer-scale growth of
  two-dimensional, phase-pure {InSe},'' \emph{Matter}, vol.~6, no.~10, pp.
  3483--3498,  2023, doi:
  \href{http://dx.doi.org/10.1016/j.matt.2023.07.012}{{10.1016/j.matt.2023.07.012}}.

\bibitem{8451970}
A.~AlMutairi and Y.~Yoon, ``Device performance assessment of monolayer hfse2: A
  new layered material compatible with high- $\kappa$ hfo2,'' \emph{IEEE
  Electron Device Letters}, vol.~39, no.~11, pp. 1772--1775,  2018, doi:
  \href{http://dx.doi.org/10.1109/LED.2018.2867957}{{10.1109/LED.2018.2867957}}.

\bibitem{yoon2011good}
Y.~Yoon, K.~Ganapathi, and S.~Salahuddin, ``How good can monolayer mos2
  transistors be?'' \emph{Nano letters}, vol.~11, no.~9, pp. 3768--3773, 2011.

\bibitem{9585026}
K.~Nandan, A.~Agarwal, S.~Bhowmick, and Y.~S. Chauhan, ``Performance
  investigation of p-fets based on highly air-stable monolayer pentagonal
  pdse2,'' \emph{IEEE Transactions on Electron Devices}, vol.~68, no.~12, pp.
  6551--6557,  2021, doi:
  \href{http://dx.doi.org/10.1109/TED.2021.3119552}{{10.1109/TED.2021.3119552}}.

\bibitem{PhysRevB.54.11169}
G.~Kresse and J.~Furthm\"uller, ``Efficient iterative schemes for ab initio
  total-energy calculations using a plane-wave basis set,'' \emph{Phys. Rev.
  B}, vol.~54, pp. 11\,169--11\,186,  Oct 1996, doi:
  \href{http://dx.doi.org/10.1103/PhysRevB.54.11169}{{10.1103/PhysRevB.54.11169}}.

\bibitem{W90}
A.~A. Mostofi, J.~R. Yates, G.~Pizzi, Y.-S. Lee, I.~Souza, D.~Vanderbilt, and
  N.~Marzari, ``An updated version of wannier90: A tool for obtaining
  maximally-localised wannier functions,'' \emph{Computer Physics
  Communications}, vol. 185, no.~8, pp. 2309--2310,  2014, doi:
  \href{http://dx.doi.org/10.1016/j.cpc.2014.05.003}{{10.1016/j.cpc.2014.05.003}}.

\bibitem{datta2005quantum}
S.~Datta, \emph{Quantum transport: atom to transistor}.\hskip 1em plus 0.5em
  minus 0.4em\relax Cambridge university press, 2005.

\bibitem{NanoTCAD}
S.~Bruzzone, G.~Iannaccone, N.~Marzari, and G.~Fiori, ``An open-source
  multiscale framework for the simulation of nanoscale devices,'' \emph{IEEE
  Transactions on Electron Devices}, vol.~61, no.~1, pp. 48--53,  2014, doi:
  \href{http://dx.doi.org/10.1109/TED.2013.2291909}{{10.1109/TED.2013.2291909}}.

\bibitem{Landauer}
M.~B\"uttiker, Y.~Imry, R.~Landauer, and S.~Pinhas, ``Generalized many-channel
  conductance formula with application to small rings,'' \emph{Phys. Rev. B},
  vol.~31, pp. 6207--6215,  May 1985, doi:
  \href{http://dx.doi.org/10.1103/PhysRevB.31.6207}{{10.1103/PhysRevB.31.6207}}.

\bibitem{shen2021ultralow}
P.-C. Shen, C.~Su, Y.~Lin, A.-S. Chou, C.-C. Cheng, J.-H. Park, M.-H. Chiu,
  A.-Y. Lu, H.-L. Tang, M.~M. Tavakoli \emph{et~al.}, ``Ultralow contact
  resistance between semimetal and monolayer semiconductors,'' \emph{Nature},
  vol. 593, no. 7858, pp. 211--217, 2021.

\bibitem{li2023approaching}
W.~Li, X.~Gong, Z.~Yu, L.~Ma, W.~Sun, S.~Gao, {\c{C}}.~K{\"o}ro{\u{g}}lu,
  W.~Wang, L.~Liu, T.~Li \emph{et~al.}, ``Approaching the quantum limit in
  two-dimensional semiconductor contacts,'' \emph{Nature}, vol. 613, no. 7943,
  pp. 274--279, 2023.

\bibitem{wu2022multiscale}
T.~Wu and J.~Guo, ``Multiscale modeling of semimetal contact to two-dimensional
  transition metal dichalcogenide semiconductor,'' \emph{Applied Physics
  Letters}, vol. 121, no.~2, 2022.

\bibitem{illarionov2020insulators}
Y.~Y. Illarionov \emph{et~al.}, ``Insulators for {2D} nanoelectronics: the gap
  to bridge,'' \emph{Nature Communications}, vol.~11, no.~1, p. 3385, 2020.

\bibitem{illarionov2019reliability}
Y.~Y. Illarionov, A.~G. Banshchikov, D.~K. Polyushkin, S.~Wachter, T.~Knobloch,
  M.~Thesberg, M.~I. Vexler, M.~Waltl, M.~Lanza, N.~S. Sokolov \emph{et~al.},
  ``Reliability of scalable mos2 fets with 2 nm crystalline caf2 insulators,''
  \emph{2D Materials}, vol.~6, no.~4, p. 045004, 2019.

\bibitem{illarionov2019ultrathin}
Y.~Y. Illarionov, A.~G. Banshchikov, D.~K. Polyushkin, S.~Wachter, T.~Knobloch,
  M.~Thesberg, L.~Mennel, M.~Paur, M.~St{\"o}ger-Pollach, A.~Steiger-Thirsfeld
  \emph{et~al.}, ``Ultrathin calcium fluoride insulators for two-dimensional
  field-effect transistors,'' \emph{Nature Electronics}, vol.~2, no.~6, pp.
  230--235, 2019.

\bibitem{5191116}
W.~Zhao \emph{et~al.}, ``Field-based capacitance modeling for sub-65-nm on-chip
  interconnect,'' \emph{IEEE Transactions on Electron Devices}, vol.~56, no.~9,
  pp. 1862--1872,  2009, doi:
  \href{http://dx.doi.org/10.1109/TED.2009.2026162}{{10.1109/TED.2009.2026162}}.

\bibitem{ma2015carrier}
N.~Ma and D.~Jena, ``Carrier statistics and quantum capacitance effects on
  mobility extraction in two-dimensional crystal semiconductor field-effect
  transistors,'' \emph{2D Materials}, vol.~2, no.~1, p. 015003, 2015.

\bibitem{1224486}
A.~Rahman, J.~Guo, S.~Datta, and M.~Lundstrom, ``Theory of ballistic
  nanotransistors,'' \emph{IEEE Transactions on Electron Devices}, vol.~50,
  no.~9, pp. 1853--1864,  2003, doi:
  \href{http://dx.doi.org/10.1109/TED.2003.815366}{{10.1109/TED.2003.815366}}.

\bibitem{9286895}
F.~Wu, H.~Tian, Z.~Yan, Y.~Shen, J.~Ren, Y.~Yang, and T.-L. Ren, ``Transistor
  subthreshold swing lowered by 2-d heterostructures,'' \emph{IEEE Transactions
  on Electron Devices}, vol.~68, no.~1, pp. 411--414,  2021, doi:
  \href{http://dx.doi.org/10.1109/TED.2020.3040350}{{10.1109/TED.2020.3040350}}.

\bibitem{doi:10.1063/1.5054281}
M.~Perucchini, E.~G.~Marin, D.~Marian, G.~Iannaccone, and G.~Fiori, ``Physical
  insights into the operation of a 1-nm gate length transistor based on mos2
  with metallic carbon nanotube gate,'' \emph{Applied Physics Letters}, vol.
  113, no.~18, p. 183507,  2018, doi:
  \href{http://dx.doi.org/10.1063/1.5054281}{{10.1063/1.5054281}}.

\bibitem{datye2022strain}
I.~M. Datye, A.~Daus, R.~W. Grady, K.~Brenner, S.~Vaziri, and E.~Pop,
  ``Strain-enhanced mobility of monolayer mos2,'' \emph{Nano Letters}, vol.~22,
  no.~20, pp. 8052--8059,  2022, doi:
  \href{http://dx.doi.org/10.1021/acs.nanolett.2c01707}{{10.1021/acs.nanolett.2c01707}}.

\bibitem{song2018largely}
C.~Song, F.~Fan, N.~Xuan, S.~Huang, G.~Zhang, C.~Wang, Z.~Sun, H.~Wu, and
  H.~Yan, ``Largely tunable band structures of few-layer {InSe} by uniaxial
  strain,'' \emph{ACS Applied Materials \& Interfaces}, vol.~10, no.~4, pp.
  3994--4000,  2018, doi:
  \href{http://dx.doi.org/10.1021/acsami.7b17247}{{10.1021/acsami.7b17247}}.

\bibitem{10.1063/5.0056448}
A.~K. Saha, M.~Si, P.~D. Ye, and S.~K. Gupta, ``{Polarization switching in
  Hf$_{0.5}$Zr$_{0.54}$O$_2$-dielectric stack: The role of dielectric layer
  thickness},'' \emph{Applied Physics Letters}, vol. 119, no.~12, p. 122903,
  09 2021, doi:
  \href{http://dx.doi.org/10.1063/5.0056448}{{10.1063/5.0056448}}.

\bibitem{cheema2022ultrathin}
S.~S. Cheema and Others, ``Ultrathin ferroic hfo2--zro2 superlattice gate stack
  for advanced transistors,'' \emph{Nature}, vol. 604, no. 7904, pp. 65--71,
  Apr 2022, doi:
  \href{http://dx.doi.org/10.1038/s41586-022-04425-6}{{10.1038/s41586-022-04425-6}}.

\bibitem{agarwal2016effect}
T.~Agarwal, I.~Radu, P.~Raghavan, G.~Fiori, A.~Thean, M.~Heyns, and W.~Dehaene,
  ``Effect of material parameters on two-dimensional materials based tfets: An
  energy-delay perspective,'' in \emph{ESSCIRC Conference 2016: 42nd European
  Solid-State Circuits Conference},  2016, doi:
  \href{http://dx.doi.org/10.1109/ESSCIRC.2016.7598241}{{10.1109/ESSCIRC.2016.7598241}}.
  pp. 55--58.

\bibitem{10091159}
A.~Naseer, K.~Nandan, A.~Agarwal, S.~Bhowmick, and Y.~S. Chauhan, ``Di-metal
  chalcogenides: A new family of promising 2-d semiconductors for
  high-performance transistors,'' \emph{IEEE Transactions on Electron Devices},
  vol.~70, no.~5, pp. 2445--2452,  2023, doi:
  \href{http://dx.doi.org/10.1109/TED.2023.3261831}{{10.1109/TED.2023.3261831}}.

\bibitem{thakar2023ultra}
K.~Thakar, B.~Rajendran, and S.~Lodha, ``Ultra-low power neuromorphic obstacle
  detection using a two-dimensional materials-based subthreshold transistor,''
  \emph{npj 2D Materials and Applications}, vol.~7, no.~1, p.~68,  Sep 2023,
  doi:
  \href{http://dx.doi.org/10.1038/s41699-023-00422-z}{{10.1038/s41699-023-00422-z}}.

\end{thebibliography}
\end{document}